# Spin-locking effect in the nanoscale ordered perovskite cobaltite LaBaCo$_2$O$_6$

Asish K. Kundu,[*] E. -L. Rautama, Ph. Boullay, V. Caignaert, V. Pralong, and

B. Raveau

CRISMAT Laboratory, 6 boulevard Maréchal Juin, Caen-14050, France

A new nanoscale ordered perovskite cobaltite, which consists of 90° ordered domains of the layered "112" LaBaCo$_2$O$_6$ has been evidenced by high resolution-transmission electron microscopy. This new form, like the disordered La$_{0.5}$Ba$_{0.5}$CoO$_3$ and ordered LaBaCo$_2$O$_6$, exhibits a ferromagnetic transition at T$_C \approx$ 179 K. However, it differs from the two previous forms by its strong magnetic anisotropy, and correlatively by its high value of coercivity (0.42 Tesla) at low temperature. We suggest that this behaviour originates from the locking of magnetic spins in the 90° oriented nano-domain. Moreover, one observes a semi-metal/semi-metal transition at T$_C$ with a maximum magnetoresistance of ~ 6.5 % at this temperature.

*Corresponding author: asish.kundu@ensicaen.fr; Fax: +33-231- 951600

Perovskite cobaltites with the general formula $Ln_{1-x}A_xCoO_{3-\delta}$ (Ln = trivalent rare earth cation, A = divalent alkaline earth cation) have been explored by many authors due to their fascinating magnetic and electron transport properties, involving complex transitions.[1-9] The latter originate from the various oxidation states of cobalt which in general exhibit mixed valent states ($Co^{2+}/Co^{3+}/Co^{4+}$), and from the various spin state configurations of $Co^{3+}$ and $Co^{4+}$ which are susceptible to change with temperature as shown for $Co^{3+}$ from low-spin (LS) $t_{2g}^6 e_g^0$, to intermediate-spin (IS) $t_{2g}^5 e_g^1$ or high-spin (HS) $t_{2g}^4 e_g^2$ states.[10-12] In those materials, the spin state configuration is mainly dictated by two competing contributions; the crystal field splitting on one hand and the exchange energy for occupied orbitals and 3d electrons on the other hand.[12,13] As a consequence, the nature of the A-site cations, especially their size, and their distribution, ordered or disordered, may influence drastically the spin state configuration of Co-ions.[5,6] In this respect, the two isochemical perovskites, $La_{0.5}Ba_{0.5}CoO_3$ and $LaBaCo_2O_6$, A-site disordered and ordered, respectively[13,14] are of great interest. Both were found to be ferromagnetic with close Curie temperatures ranging from 175 K to 190 K. However, there remains a controversy about the transport properties of the disordered perovskite $La_{0.5}Ba_{0.5}CoO_3$. Nakajima et al.[14] observed for this phase, a metallic behaviour down to 140 K, followed by an abrupt increase of resistivity below this temperature. Whereas Fauth et al.[13] observed a semi-metallic behaviour down to 110 K, but with a sudden decrease of resistivity at $T_C$. As a consequence, the first authors observed magnetoresistance (MR) below 140 K, whereas the second authors reported a maximum of MR around $T_C$.

We have revisited the perovskite with the composition "$La_{0.5}Ba_{0.5}CoO_3$" using a different method of synthesis. We report herein on a third form of this stoichiometric perovskite, that we call nanoscale ordered $LaBaCo_2O_6$, which consists of 90° oriented domains of the ordered 1:1 LaBaCo$_2$O$_6$ perovskite. We show that this ferromagnetic perovskite exhibits a T$_C$ close to the other two phases, but differs from the latter by a strong magnetic anisotropy and a much higher coercive field of 0.42 Tesla (T). Such a difference is interpreted by a spin-locking effect due to its particular nanostructure. Moreover, we observe a transition from a semi-metallic paramagnetic (PM) to ferromagnetic (FM) semi-metallic state near T$_C$ with an upturn in the resistivity below 35 K. At low temperature, the MR is only ~ 4%, whereas the highest value is observed around 179 K (~ 6.5%).

The nanostructurally ordered polycrystalline LaBaCo$_2$O$_6$ compound was synthesised by sol-gel method from metal nitrates using citric acid as a complexant. The resulting gel was then decomposed at elevated temperatures, pressed into pellets and finally sintered in Ar atmosphere at 1423 K for 48 h. For preparation of the cubic disordered La$_{0.5}$Ba$_{0.5}$CoO$_3$, the method described in Ref. 14 was applied. To obtain the fully stoichiometric materials, both of the prepared compounds were oxygenated under elevated O$_2$ pressure (130 bar) at 623 K. Oxygen content was verified by wet-chemical redox analysis to be 2.99 and 6.01 for disordered and nanostructurally ordered compounds, respectively.

The XRPD patterns were registered with a Philips X'pert Pro-diffractometer, employing Cu-K$\alpha$ radiation. Both compounds can be refined using the *Pm-3m* space group[13] with a cell parameter of $a_p$ = 0.3886(1) *nm* for disordered La$_{0.5}$Ba$_{0.5}$CoO$_3$ and $a_p$ = 0.3885(1) *nm* for the nano-ordered LaBaCo$_2$O$_6$, respectively. However, a clear *hkl*-dependent peak broadening can be detected in the case of the nanostructured LaBaCo$_2$O$_6$. The transmission electron microscopy (TEM) and high resolution electron microscopy (HREM) was carried out with a JEOL 2010F electron microscope. HREM observations at room temperature revealed that this sample is actually made of 90° oriented domains of ordered $LaBaCo_2O_6$ fitted into each other at a nanoscale level (Fig. 1). As observed by SAED on all $<100>_p$ zone axis patterns, the Fourier transform (FT) of the area (inset Fig.1) indicates clearly the existence of two sets of supercell reflections 90° oriented and compatible with the $a_p$ x $a_p$ x 2 $a_p$ supercell characteristic of the 112-type ordered structure.[8]

Temperature dependent zero field cooled (ZFC) and field cooled (FC) magnetization measured with a SQUID magnetometer (PPMS Quantum Design) in an applied field of 100 Oe is shown in Fig. 2. The curves exhibit a completely different magnetic behaviour in the same applied field conditions. A sharp increase of the magnetization occurs in the FC data around 179 K, indicating that the system is magnetically ordered and attains a spontaneous magnetization. With decreasing temperature the magnetization value increases rather slowly but does not saturate. Just below the FM transition, $T_C$ (179 K; calculated from the minimum position of the $dM_{FC}/dT$ vs temperature plot), a strong irreversibility between the ZFC and FC data appears. The ZFC curve shows a sharp peak at $T_C$ and below this, the magnetization value decreases rapidly to a lower value. By increasing the applied magnetic field (H=1 and 5 kOe, not shown), the peak in the ZFC curve broadens and shifts toward lower temperatures. The irreversible temperature between ZFC-FC, $T_{irr}$, also decreases and for a lower field $T_{irr} \sim T_C$; whereas for higher applied fields $T_{irr} < T_C$ (e.g. $T_{irr} \sim 100$ K for H=5kOe).

The strong magnetic anisotropy of this nanostructured form is also supported by the field variation of isotherm magnetization, M-H, at four different temperatures (Fig. 3). A large hysteresis loop manifests at 10 K, with a remanent magnetization ($M_r$) value of ~ 2.3 $\mu_B$/f.u. and a coercive field ($H_C$) of ~ 0.42 Tesla (T). The highest value of the magnetic moment is only ~ 3.4 $\mu_B$/f.u, which is less than the theoretical spin-only value of Co-ions in HS state. At low temperature, the relatively higher value of $H_C$ signifies a typical hard ferromagnet and with increasing temperature $H_C$ decreases gradually. Finally at temperatures higher than $T_C$, the M-H behaviour is linear corresponding to a PM state. The obtained $M_r$ and $H_C$ values of this nanostructure ordered perovskite are much higher than those previously observed for the disordered perovskite $La_{0.5}Ba_{0.5}CoO_3$, i.e. 1.7 $\mu_B$/f.u. and 0.08T respectively (inset Fig.3), as well as for the ordered perovskite $LaBaCo_2O_6$.[14] Nevertheless, the magnetic moment of the present perovskite is significantly smaller than that of the disordered form. To the best of our knowledge this is the first observation of a nanostructure ordered perovskite cobaltite with a completely different magnetic behaviour mainly due to the 90° oriented nanoscale domains.

Thus, the large divergence between the ZFC and FC magnetization at low-fields and the high coercivity can be explained in the scenario of short-range FM ordering in 90° oriented nanostructured domains extended all over the material. During the ZFC magnetization, the magnetic spins will lock or freeze in the 90° oriented nano-domains. Therefore, at low temperature the applied magnetic field is not sufficient to align the spins in its direction and it may be due to magnetostatic coupling. But, during the FC process, the spins are forced to align in the direction of the applied field at room temperature. Hence, ideally there will be no disorderliness in the spin orientation below $T_C$ as reflected in the observed value of a spontaneous magnetization. Due to this, a strong magnetic anisotropy appears in the ZFC and FC data, correlatively the pinning close to the boundaries of the nano-domains will increase the $H_C$ value.

To characterize the magnetic interaction in the PM region, we have plotted inverse of magnetic susceptibility with the variation of temperature in the 50-400 K range (inset Fig. 2). The data follow the Curie-Weiss behaviour (for T > 250 K) and a linear fit to the Curie-Weiss law yields a PM Weiss temperature ($\theta_p$) of ~ 198 K and an effective magnetic moment ($\mu_{eff}$) of ~ 5.34 $\mu_B$/f.u. The obtained value of $\theta_p$ is higher than the $T_C$ (179 K) value and the positive value of $\theta_p$ is consistent with FM interactions in the high temperature region.

The FM transition temperature is close to that of the disordered and ordered phases, which was explained by positive double exchange interactions between $Co^{3+}$ and $Co^{4+}$ ions.[13,14] The low temperature FM phase is described by the latter authors as a cluster glass but there is no supportive information. Therefore, we have studied this low temperature phase in details to understand the nature of FM ordering or to establish the cluster glass behaviour. The lack of magnetic saturation below the transition temperature may be due to short-range FM ordering. This behaviour is similar to the magnetically frustrated or glassy FM cobaltite reported in the literature.[15] This is supported by the M-H curve at higher fields (Fig. 3).

To establish this behaviour we have carried out ac-susceptibility measurements below Curie temperature ($T_C$), at different frequencies. Fig. 4 shows the in-phase, $\chi'(T)$, component of the magnetic ac-susceptibility measured at four different frequencies. The $\chi'(T)$ data is similar to the low field ZFC magnetization curve, which shows a sharp peak around 179 K. A distinct peak appears corresponding to FM ordering and, below this temperature, a weak frequency-dependent behaviour is observed and the out-of-phase component, $\chi''(T)$, also confirms this feature (inset Fig 4). Below the peak position, the magnetization value decreases with an increase in frequency. The magnetic ac-susceptibility behaviour observed for nanostructure ordered $LaBaCo_2O_6$ is quite similar to glassy-FM materials.[15]

Temperature dependence of electrical resistivity (ρ) in the presence (7T) and absence (0T) of applied magnetic field for our nanostructure ordered LaBaCo$_2$O$_6$ is shown in Fig. 5. The zero field resistivity curve shows a clear similarity with that of SrRuO$_3$[16] and SrRu$_{1-x}$M$_x$O$_3$[17] perovskites, attesting to the metallic behaviour of these oxides in the PM as well as in the FM region. With a change in slope (dρ/dT) at T$_C$, the conductivity increases more rapidly. Similar to the ruthenates, one also observes an upturn in the resistivity at low temperature, i.e. below 35 K. This was explained in SrRuO$_3$ as a weak localization contribution associated with an electron-electron interaction,[18] whereas it was interpreted as a partial orbital ordering for the ordered LaBaCo$_2$O$_6$.[14] Another explanation is that it may be due to the grain boundary effects at the domain walls. Further investigations will be needed to understand this effect.

The material is FM and semi-metallic below T$_C$ as revealed from the magnetization and electrical resistivity results. The inset in Fig. 5 depicts the magnetic field dependence on isotherm resistivity behaviour at four different temperatures (selected from the magnetic transitions). The MR is calculated as, MR(%)=[{ρ(7)-ρ(0)}/ρ(0)]x100, where ρ(0) is the sample resistivity at 0T and ρ(7) in an applied field of 7 T. The highest value of MR is observed near T$_C$ (~6.5%) like in the ruthenates[17] and the corresponding value at 10 K is only 4 %. Interestingly, the field dependent isotherm resistivity behaviour at 10 K exhibits an anisotropic effect similar to those of magnetization behaviour (see Fig. 3), also present in the 50 K isotherm resistivity data (not shown here). Also, near or above the FM transition, the irreversibility nature in the resistivity almost disappears. The low temperature isotherm MR data exhibit hysteresis effects, which resemble the "butterfly-like" feature, MR being approximately isotropic for temperatures well above the T$_C$ (at 225 and 300 K). Hence, the butterfly-like feature appears only at low temperatures (studied at 10 and 50 K) with a smaller value of MR compared to 179 K. The occurrences of anisotropic MR behaviour nearly at similar temperatures as those for isotherm M-H studies suggest the strongly correlated nature of field-induced magnetic and electronic transitions.

In conclusion, the present study demonstrates a high magnetic anisotropy in the nanoscale ordered perovskite LaBaCo$_2$O$_6$, in contrast to the disordered and ordered phases.[13,14] Our results can be explained by the locking of the cobalt spins, due to the existence of 90° oriented nanostructure domains. Both the enhanced bifurcation between the ZFC and FC magnetization and the weak frequency dependence just below T$_C$ indicate the glassy-FM or cluster glass behaviour, which signifies the competition between FM and antiferromagnetic (AFM) interactions. This suggests that the system is phase separated into large FM domains within the AFM matrix, due to the presence of both Co$^{3+}$ and Co$^{4+}$ ions.[15] The unsaturated values of M-H measurements also support this assumption. The derived effective magnetic moment (~5.34 µ$_B$/f.u.) from the experimental data corresponds to a situation where Co$^{3+}$ and Co$^{4+}$ are in LS and IS states, respectively. The obtained higher values of magnetic moment for the disordered phase compared to nanoscale ordered phase may be due to the crystal field splitting.

The authors gratefully acknowledge the CNRS and the Minister of Education and Research for financial support through their Research, Strategic and Scholarship. E.-L.R. thanks the Magnus Ehrnrooth foundation for their financial support.

**Figure captions**

**FIG. 1.** HREM image corresponding to one of the perovskite {100} planes showing the nanostructurally ordered LaBaCo$_2$O$_6$ has 90° oriented domains in the 112-type ordered phase. The inset figure shows Fourier transform (FT) of the main figure and the strong reflections related to the perovskite subcell. Two supplementary sets of reflections are observed and related to the existence of 90° oriented domains having the $a_p$x$2a_p$ 112-type ordered structure.

**FIG. 2.** Temperature dependence ZFC (open symbol) and FC (solid symbol) magnetization, M, of LaBaCo$_2$O$_6$ (H = 100 Oe) and the schematic diagrams show the existence of magnetic spins in the 90° oriented nano-domains during ZFC (below and above T$_C$). The inset figure shows inverse susceptibility, $\chi^{-1}$, vs temperature plot.

**FIG. 3.** (Color online) Field variation of magnetization (M-H) for LaBaCo$_2$O$_6$ at four different temperatures. The inset figure shows typical hysteresis curves at two different temperatures for completely disorder La$_{0.5}$Ba$_{0.5}$CoO$_3$ phase.

**FIG. 4.** (Color online) The temperature dependence in-phase component of ac-susceptibility, $\chi'$, of LaBaCo$_2$O$_6$ at four different frequencies (h$_{ac}$ = 10 Oe) and corresponding out of phase component, $\chi''$(T), data is shown in inset figure.

**FIG. 5.** (Color online) Temperature dependence of electrical resistivity, ρ, of LaBaCo$_2$O$_6$ in the presence (solid symbol) and absence (open symbol) of magnetic field (7 T). The inset figure shows magnetoresistance, MR (%), at four different temperatures.

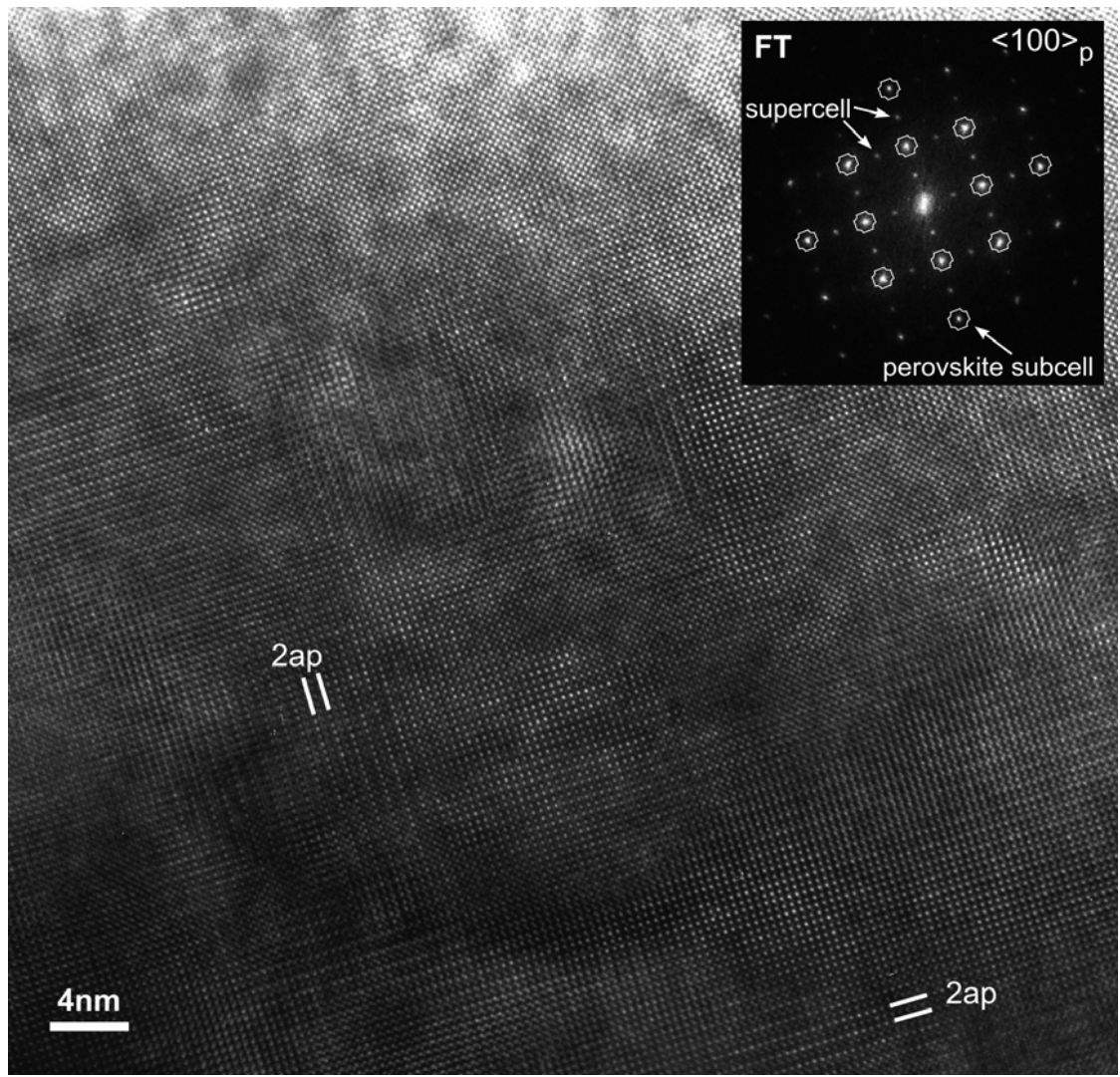

**FIG. 1.** HREM image corresponding to one of the perovskite {100} planes showing the nanostructurally ordered $LaBaCo_2O_6$ has 90° oriented domains in the 112-type ordered phase. The inset figure shows Fourier transform (FT) of the main figure and the strong reflections related to the perovskite subcell. Two supplementary sets of reflections are observed and related to the existence of 90° oriented domains having the $a_p \times 2a_p$ 112-type ordered structure.

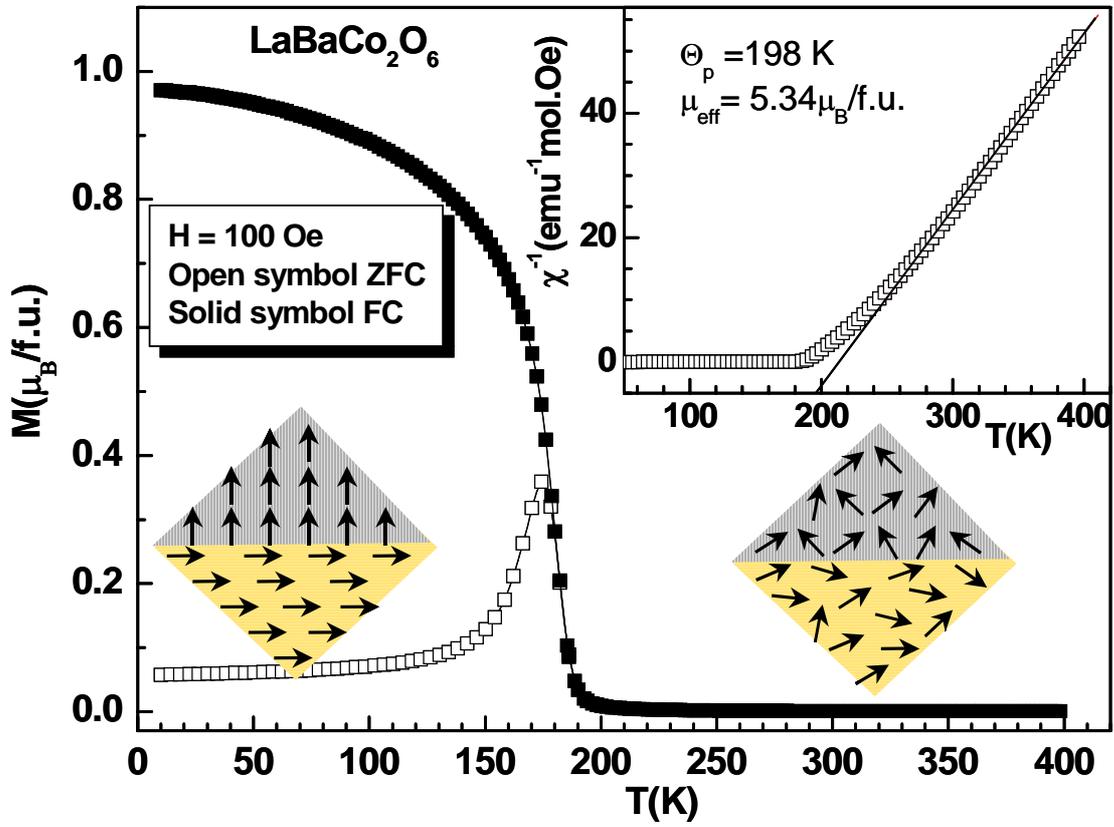

**FIG. 2.** Temperature dependence ZFC (open symbol) and FC (solid symbol) magnetization, M, of LaBaCo$_2$O$_6$ (H = 100 Oe) and the schematic diagrams show the existence of magnetic spins in the 90° oriented nano-domains during ZFC (below and above T$_C$). The inset figure shows inverse susceptibility, $\chi^{-1}$, vs temperature plot.

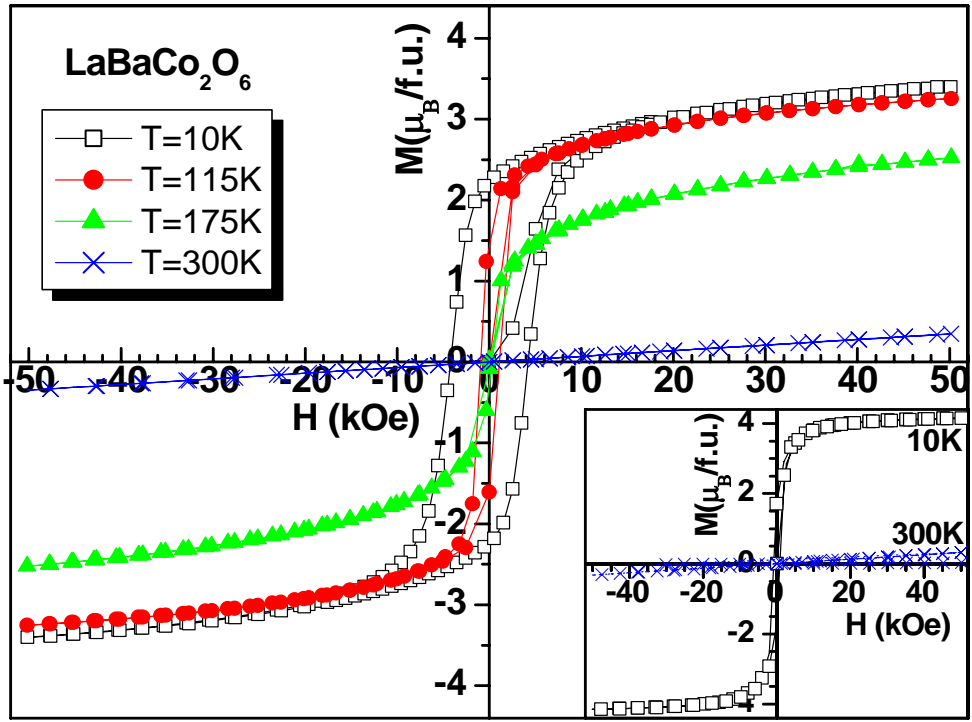

**FIG. 3.** (Color online) Field variation of magnetization (M-H) for $LaBaCo_2O_6$ at four different temperatures. The inset figure shows typical hysteresis curves at two different temperatures for completely disorder $La_{0.5}Ba_{0.5}CoO_3$ phase.

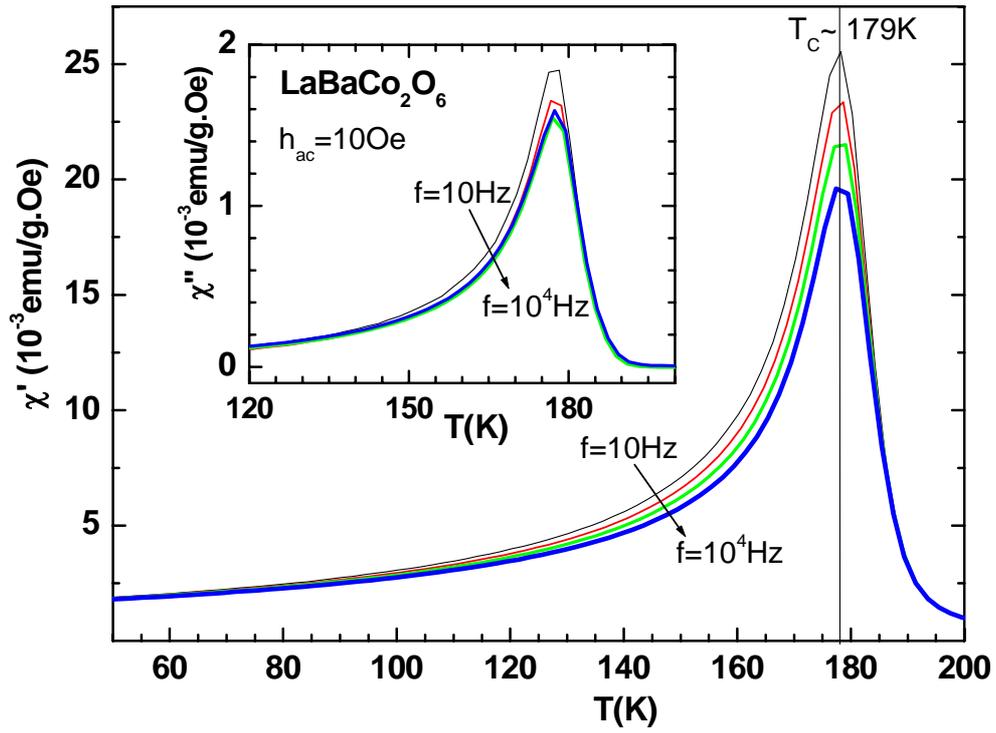

**FIG. 4.** (Color online) The temperature dependence in-phase component of ac-susceptibility, $\chi'$, of $LaBaCo_2O_6$ at four different frequencies ($h_{ac}$ = 10 Oe) and corresponding out-of-phase component, $\chi''(T)$, data is shown in inset figure.

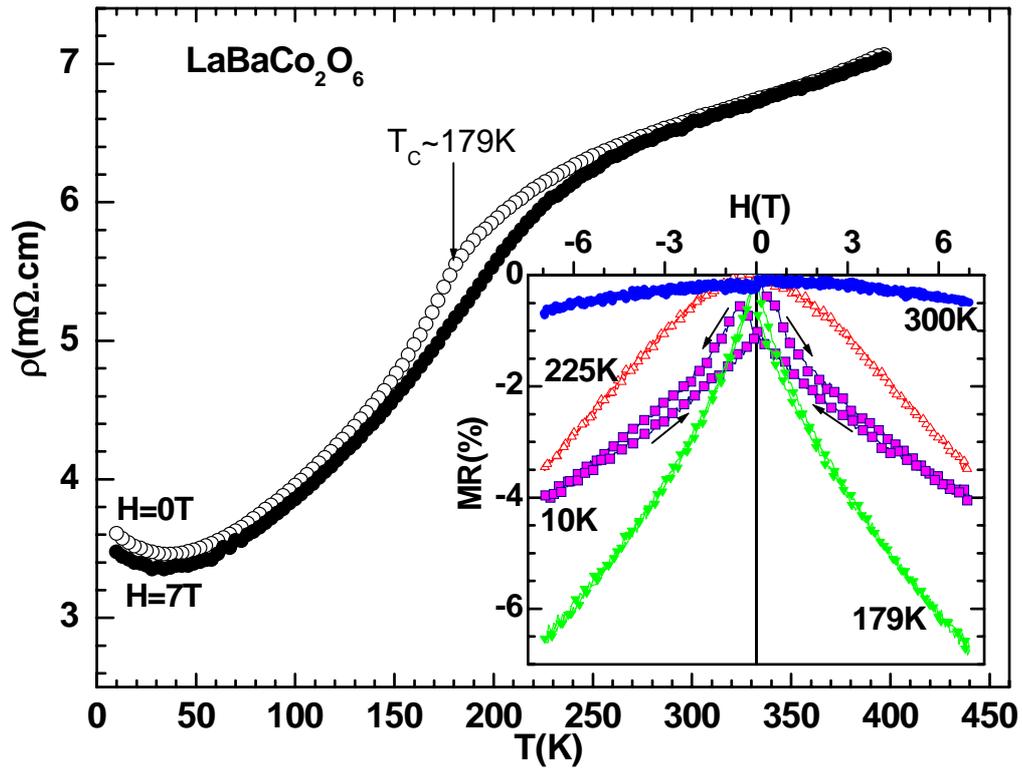

**FIG. 5.** (Color online) Temperature dependence of electrical resistivity, ρ, of LaBaCo$_2$O$_6$ in the presence (solid symbol) and absence (open symbol) of magnetic field (7 Tesla). The inset figure shows magnetoresistance, MR, at four different temperatures.